\documentclass[aps,groupedaddress,showpacs]{revtex4}
\usepackage{graphicx}
\usepackage{color}
\usepackage{amsmath}
\usepackage{bm}

\begin{document}

\title{
Linear irreversible heat engines based on the local equilibrium assumptions
}

\author{Yuki Izumida}
\thanks{Present address: Department of Complex Systems Science, Graduate School of Information Science, Nagoya University, Nagoya 464-8601, Japan}
\author{Koji Okuda}
\affiliation{Department of Information Sciences, 
Ochanomizu University, Tokyo 112-8610, Japan\\
Division of Physics, Hokkaido University, Sapporo 060-0810, Japan}

\begin{abstract}
We formulate an endoreversible finite-time Carnot cycle model based on the assumptions of local equilibrium and constant energy flux, where
the efficiency and the power are expressed in terms of the thermodynamic variables of the working substance.
By analyzing the entropy production rate caused by the heat transfer in each isothermal process during the cycle, and using an endoreversible condition applied to the linear response regime, 
we identify the thermodynamic flux and force of the present system and obtain a linear relation that connects them.
We calculate the efficiency at maximum power in the linear response regime 
by using the linear relation, which agrees with the Curzon-Ahlborn efficiency known as the upper bound in this regime.
This reason is also elucidated by rewriting our model into the form of the Onsager relations, where our model turns out to 
satisfy the tight-coupling condition leading to the Curzon-Ahlborn efficiency.
\end{abstract}

\maketitle
\section{Introduction}

The physics of heat engines originates from the Carnot's great discovery of the fundamental upper bound of the thermodynamic efficiency $\eta$
of heat engines working between two heat reservoirs with temperatures $T_h^{\rm R}$ and $T_c^{\rm R}$ ($T_h^{\rm R}>T_c^{\rm R}$)~\cite{C,REIF,MSBA}:
\begin{eqnarray}
\eta \le 1-\frac{T_c^{\rm R}}{T_h^{\rm R}}\equiv \eta_{\rm C} \ ({\rm Carnot \ efficiency})\label{eq.carnot},
\end{eqnarray}
where the equality holds for an infinitely slow process (quasistatic limit) with zero dissipation realized in, e.g., the Carnot cycle.
Indeed, this discovery may be regarded as the origin of thermodynamics itself. 
However, the quasistatic limit is an ideal case, and the thermodynamic processes observed in daily life occur at {\it finite rates}.
Remembering that we always demand power for our use of electric devices, which may originally be generated from power plants converting heat flux into electric power,
we require the physics of powerful heat engines free from the limitation of the equilibrium thermodynamics.
This deep understanding of powerful heat engines is becoming more important due to the worldwide energy crisis and climate change.

The physics of heat engines maximizing the power rather than the efficiency was 
developed in a classical paper by Curzon and Ahlborn~\cite{CA}.
They showed that, under an assumption of the endoreversible condition and the Fourier law of heat conduction between the working substance and the heat reservoir, the efficiency at maximum power $\eta^*$ of a finite-time Carnot cycle is given by
the following Curzon-Ahlborn (CA) efficiency:
\begin{eqnarray}
\eta^*=1-\sqrt{\frac{T_c^{\rm R}}{T_h^{\rm R}}}\equiv \eta_{\rm CA}=\frac{\Delta T^{\rm R}}{2T^{\rm R}}-\frac{{\Delta T^{\rm R}}^2}{8{T^{\rm R}}^2}+\cdots,\label{eq.ca}
\end{eqnarray}
where we have defined the temperature difference between the heat reservoirs as $\Delta T^{\rm R}\equiv T_h^{\rm R}-T_c^{\rm R}$ and 
the averaged temperatures between the heat reservoirs as $T^{\rm R}\equiv \frac{T_h^{\rm R}+T_c^{\rm R}}{2}$.
We note that the same formula was derived more previously in, e.g.,~\cite{N,Y} in 1950s, and it is claimed that its origin even goes back to a paper in 1920s according to a recent work~\cite{VLF}.
Because this CA efficiency displays a similar simplicity to the Carnot efficiency,
it led to the development of a new discipline of finite-time thermodynamics that aims to account for the efficiency of actual power plants and thermal devices~\cite{R,R2,LL,G,B,SNSAL,BKSST}.
The key to the derivation of the CA efficiency is the phenomenological assumption of the endoreversible condition (as named by Rubin~\cite{R} later), 
which means that the irreversibility occurs only by the heat transfer process between the working substance and the heat reservoir, and that the state of the 
working substance is internally reversible whose entropy change along the cycle is expressed by a Clausius-like equality (see Eq.~(\ref{eq.endoreversibility})).
Under this condition, the efficiency of the finite-time Carnot cycle is given by the Carnot efficiency using the ratio of the temperatures of the working substance (see Eq.~(\ref{eq.endo_carnot})), and 
the efficiency at maximum power $\eta^*$ is expressed by using the square root of the temperature ratio
$T_c^{\rm R}/T_h^{\rm R}$ as in Eq.~(\ref{eq.ca})~\cite{CA}.

Despite its importance, even until recently, there has been no argument showing whether the CA efficiency is universal as $\eta^*$ from the viewpoint of fundamental physics.
The role of the CA efficiency has become increasingly important
after Van den Broeck~\cite{VB} proved that the CA efficiency is the upper bound of $\eta^*$ in the linear response regime by
using the Onsager relations of the linear irreversible thermodynamics framework~\cite{O,DM,kc}:
\begin{eqnarray}
\eta^*\le \frac{\Delta T^{\rm R}}{2T^{\rm R}}=\eta_{\rm CA}+O\left({\Delta T^{\rm R}}^2\right),\label{eq.ca.bound}
\end{eqnarray}
where the bound is realized under the tight-coupling (no heat-leakage) condition~\cite{VB}. Because the Carnot efficiency Eq.~(\ref{eq.carnot}) can be approximated as $\eta_{\rm C}\simeq \frac{\Delta T^{\rm R}}{T^{\rm R}}$ in the linear response regime, 
it is also equivalent to say that $\eta^*$ in the linear response regime is bounded from above by one half of the Carnot efficiency as $\eta^*\le \frac{\eta_{\rm C}}{2}$.
As reviewed in~\cite{VB3,S2,T2,BCPS}, since the paper by Van den Broeck~\cite{VB}, various studies on finite-time heat engines have been conducted, which include linear response~\cite{CH,CH2,M,BSC,IO2009,IO2010,BSS,IO2014}, nonlinear response~\cite{MLVB,GMS,EKLVB,S,IO2012,WT,THR,VB2,ST,ST2}, stochastic~\cite{SS,T,VEWVB}, quantum~\cite{A,EKLVB2,BJM,AJM}, thermoelectric~\cite{GSZMS,AQGL,AQOGL}, photoelectric~\cite{REC}, molecular dynamics~\cite{IO2008,IN,HNE,SH}, and experimental~\cite{BB,ARJDSSL,RASSL} studies.

In~\cite{VB}, Van den Broeck established a view that the process of heat energy conversion into work in the linear response regime is ruled by a cross effect based on 
the Onsager relations:
\begin{eqnarray}
J_1=L_{11}X_1+L_{12}X_2,\label{eq.onsager1}\\
J_2=L_{21}X_1+L_{22}X_2,\label{eq.onsager2}
\end{eqnarray}
where $X_1$ is an ``external" thermodynamic force and $X_2$ is a ``thermal" thermodynamic force that is proportional to $\Delta T^{\rm R}$, $J_1$ and $J_2$ are their conjugate thermodynamic fluxes, and $L_{ij}$'s are the Onsager coefficients with 
reciprocity $L_{12}=L_{21}$ (see Sec.~II C for details).
While this viewpoint is familiar in steady-state heat energy conversion, such as in thermoelectric devices usually analyzed with the Onsager relations~\cite{C,BCPS,AQGL,AQOGL}, an identical formulation has also 
been established even for cyclic heat engines such as a finite-time Carnot cycle~\cite{IO2009,IO2010}.

Despite these successes, these theories of finite-time heat engines may still be abstract compared to the theory of the quasistatic heat engine.
One reason could be that a general state of a finite-time heat engine cannot be drawn on a thermodynamic plane, 
thus a clear picture is lacking, unlike the quasistatic cycle with well-defined thermodynamic variables of the working substance such as the pressure, volume, and so on.
Even in a finite-time cycle, however, it may still be possible to assume 
that the state of the working substance is specified by a {\it unique} combination of the thermodynamic variables at any instant along the cycle, 
and hence the working substance and the heat reservoirs are in a {\it local} equilibrium, but not in a global equilibrium similar to the quasistatic cycle.
According to this local equilibrium assumption, we can draw the finite-time cycle on the thermodynamic plane as well as the quasistatic cycle, 
and can also assume that the fundamental thermodynamic relations hold between the thermodynamic variables of the working substance during a finite-time cycle.
In fact, Rubin introduced this local equilibrium thermodynamic description to the endoreversible cycle very previously~\cite{R} (see also~\cite{WT}),
where the endoreversible condition is expressed by the entropy change of the working substance during one cycle.
However, this local equilibrium assumption has not fully been taken into account in the recent literatures, and 
we hence are naturally motivated to elucidate how these endoreversible heat engine models based on the local equilibrium assumption are unified with the more recent linear irreversible thermodynamic description using the Onsager relations Eqs.~(\ref{eq.onsager1}) and (\ref{eq.onsager2}).

In the present study, we formulate an endoreversible finite-time Carnot cycle model based on the assumptions of local equilibrium and constant energy flux.
In our framework, the power and the efficiency can be expressed in terms of the thermodynamic variables of the working substance.
From the analysis of the entropy production rate caused by the heat transfer in each isothermal process during the cycle, 
we identify the thermodynamic flux and force in each isothermal process, where the flux and force are assumed to be related by the Fourier law.
We also find that, due to the endoreversible condition applied to the linear response regime, 
these thermodynamic forces in the isothermal processes are not independent, and we can identify the reduced thermodynamic force, whose conjugate thermodynamic flux turns out be an averaged heat flux.
We then obtain a linear relation 
that connects these thermodynamic flux and force.
From the calculation of the efficiency at maximum power by using this linear relation, 
we obtain the CA efficiency as $\eta^*$ of the present model.
Then, suitably changing the variables, we elucidate that the linear relation in our framework can be rewritten into the form of the Onsager relations, 
from which we can directly confirm the tight-coupling condition leading to the CA efficiency in the framework of~\cite{VB}.
Therefore our work establishes a precise connection between the finite-time thermodynamic approach to heat engines by Curzon and Ahlborn 
and the linear irreversible thermodynamic approach based on the Onsager relations~\cite{VB} via the local equilibrium assumption that gives us a more intuitive picture of the finite-time heat engines.

The present paper is organized as follows. In Sec.~II A, we introduce our model based on the assumptions of local equilibrium and constant energy flux.
In Sec.~II B, by analyzing the entropy production rate of our heat engine with the aid of the endoreversible condition applied to the linear response regime, we naturally identify the thermodynamic flux and force and a linear relation that connects them. 
We then calculate the efficiency at maximum power in the linear response regime by using this linear relation.
In Sec.~II C, we elucidate the relationship between the linear relation obtained in Sec.~II B and the Onsager relations Eqs.~(\ref{eq.onsager1}) and (\ref{eq.onsager2}), explicitly showing that our model surely satisfies the tight-coupling (no heat-leakage) condition leading to the CA efficiency. In Sec.~III, we discuss a few aspects related to our formulation in Sec.~II, 
and summarize our study.

\section{Model and results}
\subsection{Local equilibrium thermodynamic formulation of the endoreversible finite-time Carnot cycle}
Our heat engine model consists of the working substance, the hot heat reservoir with temperature $T_h^{\rm R}$ and the cold heat reservoir with  temperature $T_c^{\rm R}$.
We assume that the working substance is always in a local equilibrium state 
specified by a unique combination of the well-defined thermodynamic variables, and 
the heat reservoirs are also in a local equilibrium state. 
Denoting the internal energy and the entropy of the heat reservoir by $U^{\rm R}_i$ and $S^{\rm R}_i$ ($i=h, c$), respectively, 
the temperature is defined by $\frac{1}{T^{\rm R}_i}\equiv \frac{\partial S^{\rm R}_i}{\partial U^{\rm R}_i}$. 
We also denote the internal energy and entropy of the working substance by $U$ and $S$, respectively.
Hereafter, we use the suffix $i$ to denote the thermodynamic variable of the working substance when it contacts with the heat reservoir with the temperature $T_i^{\rm R}$.  
We then obtain the first law of thermodynamics (energy-conservation law) by using these thermodynamic variables as~\cite{R}
\begin{eqnarray}
-\frac{dU_i^{\rm R}}{dt}=\frac{dU_i}{dt}+P_i \frac{dV_i}{dt},\label{eq.JQ}
\end{eqnarray}
where we define $\frac{1}{T_i}\equiv \frac{\partial S_i}{\partial U_i}$ and $\frac{P_i}{T_i}\equiv \frac{\partial S_i}{\partial V_i}$
through the fundamental thermodynamic relation 
\begin{eqnarray}
dU_i(S_i,V_i)=\frac{\partial U_i}{\partial S_i}dS_i+\frac{\partial U_i}{\partial V_i}dV_i=T_idS_i-P_idV_i,\label{eq.fundamental}
\end{eqnarray}
with $T_i$, $P_i$, and $V_i$ being the temperature, pressure, and volume of the working substance, respectively.
Eq.~(\ref{eq.JQ}) states that the heat flux from the heat reservoir $-\frac{dU_i^{\rm R}}{dt}$, which is 
the internal-energy change rate of the heat reservoir, is decomposed into the internal-energy change 
rate of the working substance $\frac{dU_i}{dt}$ and the instantaneous power output $P_i\frac{dV_i}{dt}$. 

\begin{figure}
\includegraphics[scale=0.45]{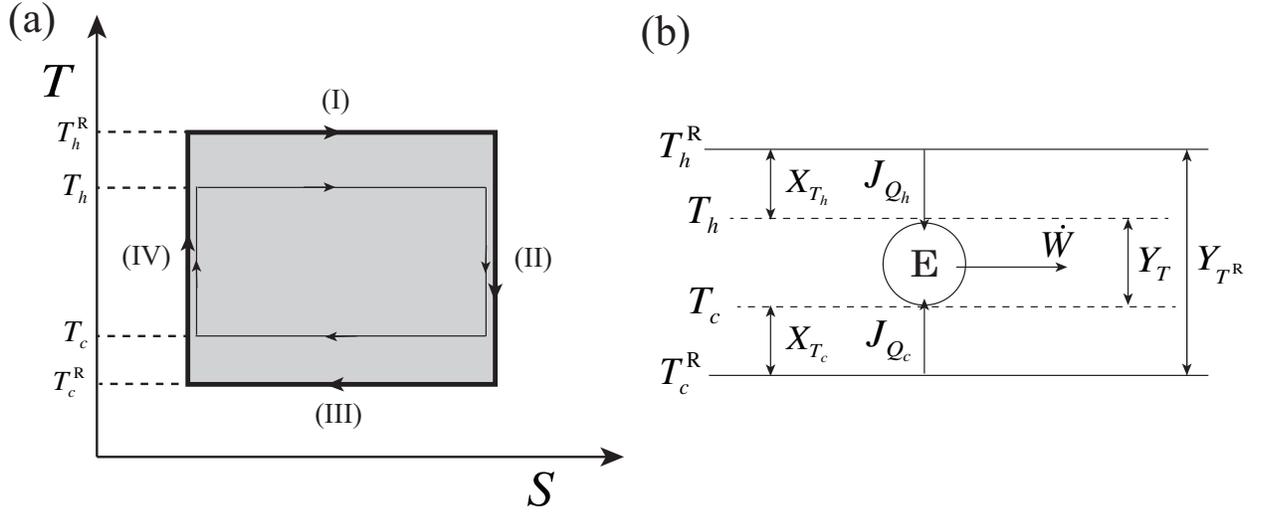}
\caption{(a) Temperature--entropy ($T$--$S$) 
diagram of an endoreversible finite-time Carnot cycle based on the assumptions of local equilibrium and constant energy flux, where $T$ and $S$ 
are well-defined (the inner cycle depicted with the thin solid curve) 
on a thermodynamic plane even far from the quasistatic limit (the large outer cycle depicted with the bold solid curve).
The cycle consists of (I) isothermal expansion process in contact with the hot heat reservoir with the temperature $T_h^{\rm R}$, (II) adiabatic expansion process, (III) isothermal compression process in contact with the cold heat reservoir with the temperature $T_c^{\rm R}$, and (IV) adiabatic compression process.
In the inner finite-time cycle, the temperature of the working substance $T_i$ during the isothermal process is assumed to be constant at any instant, but 
does not agree with $T_i^{\rm R}$. 
The area of the cycle represents the work output $W$ during one cycle: $W=\Delta T^{\rm R}\Delta S$ 
for the quaistatic cycle and $W=\Delta T\Delta S$ for the finite-time cycle.
(b) Schematic illustration of the heat energy conversion using the endoreversible heat engine model, where the working substance specified by a 
unique combination of the thermodynamic variables and the heat reservoirs are assumed to be in a local equilibrium state.
The thermodynamic fluxes and forces in the figure are defined in~Eqs.~(\ref{eq.def_flux_force}), (\ref{eq.JQYTR}) and (\ref{eq.JSYT}).
}\label{engine}
\end{figure}

Our heat engine experiences a thermodynamic cycle that consists of (I) an isothermal expansion process in contact with the hot heat reservoir with the temperature $T_h^{\rm R}$, (II) an adiabatic expansion process, (III) an isothermal compression process in contact with the cold heat reservoir with the temperature $T_c^{\rm R}$, and (IV) an adiabatic compression process (see Fig.~\ref{engine} (a)).
We assume that we can take the durations of the adiabatic processes to be sufficiently short compared to the ones of the isothermal processes in a finite-time cycle, and that the thermodynamic states of the working substance move along the quasistatic adiabatic curves in the thermodynamic plane, while the thermodynamic states of the working substance during the isothermal processes do not agree with the quasistatic isothermal curves. 
This assumption is quite natural because relaxation to the equilibrium state during the adiabatic process is occurred by internal processes inside the working substance itself, which 
is much faster than the speed of relaxation to the (global) equilibrium during the isothermal process determined by interactions with the heat reservoir.
From this, our heat engine can be regarded to run one cycle in the cycle time $t_{\rm cyc} \equiv t_h+t_c$, where we denote by $t_i$ the duration of the isothermal process in contact with the heat reservoir at $T_i^{\rm R}$.
Additionally, in our formulation, we also assume that the energy flux corresponding to each term in Eq.~(\ref{eq.JQ}) is constant at any instant along each isothermal process, and the temperature $T_i$ does not change during the isothermal process, where these assumptions are also adopted in the original CA model~\cite{CA} (see Sec.~III for a discussion on this assumption). 
Using Eq.~(\ref{eq.fundamental}), we can rewrite Eq.~(\ref{eq.JQ}) as
\begin{eqnarray}
-\frac{dU_i^{\rm R}}{dt}=T_i\frac{dS_i}{dt}.\label{eq.JQ_S1}
\end{eqnarray}
The heat from the heat reservoir during the isothermal process $Q_i$ is calculated by using Eq.~(\ref{eq.JQ_S1}) as
\begin{eqnarray}
&&Q_h\equiv -\int_0^{t_h}\frac{dU_h^{\rm R}}{dt}dt=T_h\Delta S_h,\label{eq.Qh}\\ 
&&Q_c\equiv -\int_{t_h}^{t_{\rm cyc}}\frac{dU_c^{\rm R}}{dt}dt=T_c\Delta S_c,\label{eq.Qc}
\end{eqnarray}
where we defined the entropy change of the working substance during the isothermal process as
\begin{eqnarray}
\Delta S_h\equiv \int_0^{t_h}\frac{dS_h}{dt}dt, \ \Delta S_c\equiv \int_{t_h}^{t_{\rm cyc}}\frac{dS_c}{dt}dt.
\end{eqnarray}
Because we require that the cycle is closed after one cycle and we also assume that the adiabatic processes are regarded as 
quasistatic processes, the following relations should hold:
\begin{eqnarray}
\Delta S_h=-\Delta S_c\equiv \Delta S.\label{eq.entropy_change}
\end{eqnarray}
We note that the endoreversibility condition~\cite{CA,R}
\begin{eqnarray}
\frac{{Q}_h}{T_h}+\frac{{Q}_c}{T_c}=0\label{eq.endoreversibility}
\end{eqnarray} 
automatically holds from Eqs.~(\ref{eq.Qh}), (\ref{eq.Qc}) and (\ref{eq.entropy_change}).
This condition manifests that the entropy production is occurred only by the heat transfer between the heat reservoirs and the working substance, and the
internal state of the working substance is, as assumed above, in a local equilibrium state, which implies that the entropy change of the working substance along the cycle is expressed by the Clausius-like equality using the temperature of the working substance. 
Eq.~(\ref{eq.endoreversibility}) also implies that the efficiency of this type of the heat engines is given by the ``endoreversible Carnot efficiency" using the temperatures of the working substance as
\begin{eqnarray}
\eta \equiv \frac{W}{Q_h}=1-\frac{T_c}{T_h}.\label{eq.endo_carnot}
\end{eqnarray} 
Although this differs from the usual Carnot efficiency Eq.~(\ref{eq.carnot}) that 
uses the temperatures of the heat reservoirs, 
it is remarkable that the efficiency is still expressed by the thermodynamic variables of the working substance.
In this way, the endoreversibility condition Eq.~(\ref{eq.endoreversibility}) largely constrains the behavior of the heat engines.
The power output $\dot{W}$ of the heat engine is also expressed by using the entropy change as
\begin{eqnarray}
\dot{W}\equiv \frac{W}{t_{\rm cyc}}=\frac{Q_h+Q_c}{t_{\rm cyc}}=\frac{\Delta T \Delta S}{t_{\rm cyc}},\label{eq.power2}
\end{eqnarray}
where $\Delta T\equiv T_h-T_c$ is the temperature difference of the working substance between the isothermal processes and we used the first law of thermodynamics $W=Q_h+Q_c$ for one cycle.
Hereafter we denote by the dot the quantity divided by the cycle period.
Then the rightmost expression in Eq.~(\ref{eq.power2}) using the ``endoreversible work" $W=\Delta T\Delta S$ written by the thermodynamic variables of the working substance like the quasistatic work $W=\Delta T^{\rm R}\Delta S$ 
may be regarded as ``endoreversible power." 
The power Eq.~(\ref{eq.power2}) and the efficiency Eq.~(\ref{eq.endo_carnot}), 
which are defined by using the thermodynamic variables of the working substance in this way,  
are a characteristic of our local equilibrium description of the endoreversible heat engine model. 
Especially, the ``endoreversible power" expression has never been used in a context of determining the efficiency at maximum power so far.
We note that these expressions also give us an intuitive picture underlying heat-energy conversions in a finite-time Carnot cycle, 
because the efficiency and the work output are still given by the quasistatic-like expressions.

\subsection{Efficiency at maximum power in the linear response regime}
In this subsection, we consider the efficiency at maximum power of our heat engine in the linear response regime $\Delta T^{\rm R}\to 0$, 
based on the local equilibrium assumption introduced in Sec.~II A.

Our analysis using the linear irreversible thermodynamics begins from the entropy production rate.
Because we assume that the heat reservoirs and the working substance are always in a local equilibrium state with the well-defined entropies, 
the entropy production rate $\frac{ds}{dt}$ of the total system (the working substance and the heat reservoirs)
at any instant along the isothermal process is expressed by the sum of the entropy change rates of these partial systems:
\begin{eqnarray}
\frac{ds_i}{dt}&&\equiv \frac{dS_i^{\rm R}}{dt}+\frac{dS_i}{dt}=\frac{\partial S_i^{\rm R}}{\partial U_i^{\rm R}}\frac{dU_i^{\rm R}}{dt}+\left(\frac{\partial S_i}{\partial U_i}\frac{dU_i}{dt}+\frac{\partial S_i}{\partial V_i}\frac{dV_i}{dt}\right)
=\left(\frac{1}{T_i^{\rm R}}-\frac{1}{T_i}\right)\frac{dU_i^{\rm R}}{dt},
\end{eqnarray}
where we used Eq.~(\ref{eq.JQ}).
Then, the entropy production rate for one cycle $\dot{\sigma}$ is written as
\begin{eqnarray}
\dot{\sigma}=\frac{1}{t_{\rm cyc}}\oint ds=\frac{1}{t_{\rm cyc}}\int_0^{t_{\rm cyc}}\frac{ds(t)}{dt}dt
&&=\frac{1}{t_{\rm cyc}}\int_0^{t_h}\left(\frac{1}{T_h^{\rm R}}-\frac{1}{T_h}\right)\frac{dU_h^{\rm R}}{dt}dt+\frac{1}{t_{\rm cyc}}\int_{t_h}^{t_{\rm cyc}}\left(\frac{1}{T_c^{\rm R}}-\frac{1}{T_c}\right)\frac{dU_c^{\rm R}}{dt}dt\nonumber\\
&&=\sum_{i}\left(\frac{1}{T_i^{\rm R}}-\frac{1}{T_i}\right)\frac{\Delta U_i^{\rm R}}{t_{\rm cyc}} \equiv \sum_i J_{Q_i}X_{T_i},\label{eq.entropy_product}
\end{eqnarray}
where we defined the internal energy change of the heat reservoir during the isothermal process as
\begin{eqnarray}
\Delta U_h^{\rm R}\equiv \int_0^{t_h}\frac{dU_h^{\rm R}}{dt}dt, \ \Delta U_c^{\rm R}\equiv \int_{t_h}^{t_{\rm cyc}}\frac{dU_c^{\rm R}}{dt}dt.
\end{eqnarray}
From Eq.~(\ref{eq.entropy_product}), we naturally define the thermodynamic force as the (inverse) temperature difference between the working substance and the heat reservoir, 
and we define the conjugate thermodynamic flux as the heat flux from the heat reservoir~\cite{WT} (see Fig.~\ref{engine} (b)):
\begin{eqnarray}
J_{Q_i}\equiv -\frac{\Delta U^{\rm R}_i}{t_{\rm cyc}}=\frac{{Q}_i}{t_{\rm cyc}}, \ X_{T_i}\equiv -\left(\frac{1}{T_i^{\rm R}}-\frac{1}{T_i}\right).\label{eq.def_flux_force}
\end{eqnarray}
Because the energy flux and the temperature of the working substance is assumed to be constant along each isothermal process, 
we obtain
\begin{eqnarray}
-\frac{\Delta U_i^{\rm R}}{t_i}=-\frac{dU_i^{\rm R}}{dt}=T_i\frac{dS_i}{dt},
\end{eqnarray}
by using Eq.~(\ref{eq.JQ_S1}).  
Then the thermodynamic flux $J_{Q_i}$ can also be expressed by using the time derivative of the thermodynamic variable of the working substance as
\begin{eqnarray}
J_{Q_i}=a_iT_i\frac{dS_i}{dt},\label{eq.JQdSdt}
\end{eqnarray}
where we denote by $a_i$ the ratio of the duration of each isothermal process $t_i$ to the cycle time $t_{\rm cyc}$.
To proceed further, we need a relation that connects $J_{Q_i}$ and $X_{T_i}$, in addition to the local equilibrium thermodynamic formulation mentioned in Sec.~II A.
Because we are adopting the local equilibrium thermodynamic assumption, it is also quite natural to assume that 
the heat flows in proportion to the temperature difference (the Fourier law) in the same way as the original CA model~\cite{CA}:
\begin{eqnarray}
Q_i=\kappa_i (T_i^{\rm R}-T_i)t_i,\label{eq.fourier}
\end{eqnarray}
where we denote by $\kappa_i$ the thermal conductance between the heat reservoir with the temperature $T_i^{\rm R}$ and the working substance. 
Using Eq.~(\ref{eq.fourier}), we then obtain the following relationship between $J_{Q_i}$ and $X_{T_i}$:
\begin{eqnarray}
J_{Q_i}=\frac{Q_i}{t_{\rm cyc}}=a_i\kappa_i(T_i^{\rm R}-T_i)=a_i\kappa_iT_i^{\rm R}T_i X_{T_i}.\label{eq.jq_fourier}
\end{eqnarray}

In the following, we consider the linear response regime $\Delta T^{\rm R}\to 0$.
By substituting Eq.~(\ref{eq.fourier}) with $a_it_{\rm cyc}$ instead of $t_i$ into the endoreversibility condition Eq.~(\ref{eq.endoreversibility}), 
and expanding that with respect to $\Delta T$ and $\Delta T^{\rm R}$, we can write Eq.~(\ref{eq.endoreversibility}) as 
\begin{eqnarray}
\left(a_h \kappa_h+a_c\kappa_c\right)\left(\frac{T^{\rm R}}{T}-1\right)+\left(a_h \kappa_h-a_c\kappa_c\right)\frac{T^{\rm R}}{2T}\left(-\frac{\Delta T}{T}+\frac{\Delta T^{\rm R}}{T^{\rm R}}\right)+O(\Delta T^{\rm R}\Delta T,\Delta T^2)=0,\label{eq.endoreversibility_2}
\end{eqnarray}
where $T\equiv \frac{T_h+T_c}{2}$ is the averaged temperature of the working substance between the isothermal processes.
Then, by substituting $T=c^{(0)}+c^{(1)}_{\rm R}\Delta T^{\rm R}+c^{(1)}\Delta T$ into Eq.~(\ref{eq.endoreversibility_2}), we obtain the expansion coefficients order by order as follows:
\begin{eqnarray}
&&c^{(0)}=T^{\rm R},\label{eq.endo_cond_1}\\
&&c^{(1)}_{\rm R}=\frac{a_h\kappa_h-a_c\kappa_c}{2(a_h\kappa_h+a_c\kappa_c)},\label{eq.endo_cond_2}\\
&&c^{(1)}=-\frac{a_h\kappa_h-a_c\kappa_c}{2(a_h\kappa_h+a_c\kappa_c)}.\label{eq.endo_cond_3}
\end{eqnarray}
Then, $T_h=T+\frac{\Delta T}{2}$ and $T_c=T-\frac{\Delta T}{2}$ are expressed as
\begin{eqnarray}
&&T_h=T^{\rm R}+\frac{a_h\kappa_h-a_c\kappa_c}{2(a_h\kappa_h+a_c\kappa_c)}\Delta T^{\rm R}+\left(\frac{1}{2}-\frac{a_h\kappa_h-a_c\kappa_c}{2(a_h\kappa_h+a_c\kappa_c)}\right)\Delta T,\label{eq.Th}\\
&&T_c=T^{\rm R}+\frac{a_h\kappa_h-a_c\kappa_c}{2(a_h\kappa_h+a_c\kappa_c)}\Delta T^{\rm R}+\left(-\frac{1}{2}-\frac{a_h\kappa_h-a_c\kappa_c}{2(a_h\kappa_h+a_c\kappa_c)}\right)\Delta T.\label{eq.Tc}
\end{eqnarray}
By using Eqs.~(\ref{eq.endo_cond_1})--(\ref{eq.endo_cond_3}), we also find that $X_{T_i}$'s are expressed in terms of $\Delta T$ and $\Delta T^{\rm R}$ as follows:
\begin{eqnarray}
X_{T_h}&&\simeq \frac{T_h^{\rm R}-T_h}{{T^{\rm R}}^2}=\frac{a_c\kappa_c}{a_h\kappa_h+a_c\kappa_c}\frac{\Delta T^{\rm R}-\Delta T}{{T^{\rm R}}^2}=\frac{a_c\kappa_c}{a_h\kappa_h+a_c\kappa_c}X_T,\label{eq.xth_approx}\\
X_{T_c}&&\simeq \frac{T_c^{\rm R}-T_c}{{T^{\rm R}}^2}=-\frac{a_h\kappa_h}{a_h\kappa_h+a_c\kappa_c}\frac{\Delta T^{\rm R}-\Delta T}{{T^{\rm R}}^2}=-\frac{a_h\kappa_h}{a_h\kappa_h+a_c\kappa_c}X_T,\label{eq.xtc_approx}
\end{eqnarray}
where we defined the ``reduced thermodynamic force" $X_T$ as
\begin{eqnarray}
X_T\equiv \frac{\Delta T^{\rm R}-\Delta T}{{T^{\rm R}}^2}.\label{eq.reduced_force}
\end{eqnarray}
From Eqs.~(\ref{eq.xth_approx}) and (\ref{eq.xtc_approx}), we find that $X_{T_h}$ and $X_{T_c}$ are proportional with each other in the linear response regime.
By using Eqs.~(\ref{eq.xth_approx}) and (\ref{eq.xtc_approx}) and approximating $J_{Q_i}$ in Eq.~(\ref{eq.jq_fourier}) as $J_{Q_i}\simeq a_i\kappa_i {T^{\rm R}}^2X_{T_i}$, we can simplify the entropy production rate Eq.~(\ref{eq.entropy_product}) by using $X_T$ up to the quadratic order of $X_T$ as
\begin{eqnarray}
\dot{\sigma}=\frac{1}{t_{\rm cyc}}\oint ds=\sum_i J_{Q_i}X_{T_i}\simeq \sum_i a_i\kappa_i {T^{\rm R}}^2 {X_{T_i}}^2=\frac{a_h\kappa_ha_c\kappa_c}{a_h\kappa_h+a_c\kappa_c}{T^{\rm R}}^2X_T^2\equiv J_QX_T,\label{eq.entropy_product_quadratic}
\end{eqnarray}
where $J_Q$ is the averaged heat flux~\cite{ST,JM,NK} defined as
\begin{eqnarray}
J_Q\equiv \frac{J_{Q_h}-J_{Q_c}}{2}=\frac{a_h\kappa_ha_c\kappa_c}{a_h\kappa_h+a_c\kappa_c}{T^{\rm R}}^2X_T.\label{eq.JQXT}
\end{eqnarray}
Therefore the description of the present heat engine model is reduced to this linear relation Eq.~(\ref{eq.JQXT}).
In the linear response regime, the endoreversible Carnot efficiency Eq.~(\ref{eq.endo_carnot}) and the endoreversible power Eq.~(\ref{eq.power2}) 
are also approximated by using $X_T$ as
\begin{eqnarray}
\eta &&\simeq \frac{\Delta T}{T^{\rm R}}=\frac{\Delta T^{\rm R}}{T^{\rm R}}-T^{\rm R}X_T,\label{eq.eta_xt}\\
\dot{W}&&=J_{Q_h}\eta \simeq J_{Q_h} \frac{\Delta T}{T^{\rm R}}=\frac{a_h\kappa_ha_c\kappa_c}{a_h\kappa_h+a_c\kappa_c}T^{\rm R}X_T\left(\Delta T^{\rm R}-{T^{\rm R}}^2X_T\right),\label{eq.pow_xt}
\end{eqnarray}
respectively, where $\dot{W}$ is a quadratic function of the reduced thermodynamic force $X_T$.
Because $X_T$ is proportional to the temperature difference between the working substance and the heat reservoir as in Eqs.~(\ref{eq.xth_approx}) and (\ref{eq.xtc_approx}), 
it is natural that we can control the power by changing $X_T$: 
 the quasistatic limit is realized under no temperature difference between them as $X_T^{\rm qs}=0$ ($\Delta T^{\rm qs}=\Delta T^{\rm R}$).
Then the heat flux $J_{Q_i}$ also vanishes from the linear Fourier law 
Eq.~(\ref{eq.jq_fourier}) via Eqs.~(\ref{eq.xth_approx}) and (\ref{eq.xtc_approx}), and we obtain no power $\dot{W}=0$ from Eq.~(\ref{eq.pow_xt}) as in the usual Carnot cycle. As $X_T$ increases from $X_T^{\rm qs}$, the heat flux $J_{Q_i}$ becomes finite.
Since the power $\dot{W}$ is a quadratic function of $X_T$, it can take a maximum value at a certain point, which is determined by $\frac{\partial \dot{W}}{\partial X_T}=0$:
\begin{eqnarray}
X_T^*=\frac{\Delta T^{\rm R}}{2{T^{\rm R}}^2} \  \left(\Delta T^*=\frac{\Delta T^{\rm R}}{2}\right).\label{eq.maxpow_condition}
\end{eqnarray}
In the quasistatic limit $X_T^{\rm qs}=0$, we attain the Carnot efficiency $\frac{\Delta T^{\rm R}}{T^{\rm R}} \simeq \eta_{\rm C}$ as the maximum efficiency from Eq.~(\ref{eq.eta_xt}).
At the maximum power, from Eqs.~(\ref{eq.eta_xt}) and (\ref{eq.maxpow_condition}), the efficiency $\eta^*$ is given by
\begin{eqnarray}
\eta^*=\frac{\Delta T^{\rm R}}{2T^{\rm R}}.\label{eq.eta_maxp_lin}
\end{eqnarray}
This is the CA efficiency up to to first order of $\Delta T^{\rm R}$ that corresponds to the equality in Eq.~(\ref{eq.ca.bound}).
The maximum power $\dot{W}^*$ is also given as
\begin{eqnarray}
\dot{W}^*=\frac{a_h \kappa_h a_c\kappa_c}{a_h\kappa_h+a_c\kappa_c} \frac{{\Delta T^{\rm R}}^2}{4{T^{\rm R}}},
\end{eqnarray}
which depends on the thermal conductivity~\cite{CA}.
Here we note that we obtain the CA efficiency in Eq.~(\ref{eq.eta_maxp_lin}) without using the Onsager relations Eqs.~(\ref{eq.onsager1}) and (\ref{eq.onsager2}), but using the linear relation between the thermodynamic flux and force Eq.~(\ref{eq.JQXT}) with the aid of the endoreversible expression of the efficiency and power Eqs.~(\ref{eq.eta_xt}) and (\ref{eq.pow_xt}). In the next Sec.~II C, we consider this connection.

\subsection{Formulation of the endoreversible finite-time Carnot cycle model using Onsager relations}
As we have shown in Sec.~II B, the efficiency at maximum power $\eta^*$ in Eq.~(\ref{eq.eta_maxp_lin}), which is based on the local equilibrium thermodynamic formulation using the linear relation Eq.~(\ref{eq.JQXT}), is the upper bound in Eq.~(\ref{eq.ca.bound}), while the inequality in Eq.~(\ref{eq.ca.bound}) comes from the formulation based on the Onsager relations~\cite{VB}.
Therefore, we elucidate the relationship between these formulations in this subsection.

First, we briefly review the derivation of the inequality for the efficiency at maximum power in Eq.~(\ref{eq.ca.bound})~\cite{VB}.
Denoting an external force and its conjugate variable by $F$ and $x$, respectively, we can generally express the power of the heat engine $\dot{W}$ as $\dot{W}=-F\dot{x}$.
Then the entropy production rate of the total system $\dot{\sigma}$ is decomposed into the sum of the entropy increase rate of each heat reservoir 
because the state of the working substance should return to the original state after one cycle:
\begin{eqnarray}
\dot{\sigma}=-\frac{\dot{Q}_h}{T_h^{\rm R}}-\frac{\dot{Q}_c}{T_c^{\rm R}}=\left(\frac{1}{T_c^{\rm R}}-\frac{1}{T_h^{\rm R}}\right)\dot{Q}_h-\frac{\dot{W}}{T_c^{\rm R}}
\simeq \frac{F\dot{x}}{T^{\rm R}}+\frac{\Delta T^{\rm R}}{{T^{\rm R}}^2}\dot{Q}_h=J_1X_1+J_2X_2.\label{eq.entropy_vb}
\end{eqnarray}
Here, the thermodynamic fluxes $J_1\equiv \dot{x}$ and $J_2\equiv \dot{Q}_h$, and their conjugate thermodynamic forces $X_1\equiv \frac{F}{T^{\rm R}}$ and $X_2\equiv \frac{\Delta T^{\rm R}}{{T^{\rm R}}^2}$ are related through the Onsager relations Eqs.~(\ref{eq.onsager1}) and (\ref{eq.onsager2}).
We note that, in this case, we do not necessarily assume that the working substance along the cycle is expressed in terms of the well-defined thermodynamic variables, in contrast to 
our formulation in Sec~II B.
Using these thermodynamic fluxes and forces,
the power and the efficiency  are given as
\begin{eqnarray}
&&\dot{W}=-J_1X_1T^{\rm R},\label{eq.ex_force}\\
&&\eta=\frac{\dot{W}}{\dot{Q}_h}=-\frac{J_1X_1T^{\rm R}}{J_2}.\label{eq.effi_lin}
\end{eqnarray}
With these expressions as well as the Onsager relations Eqs.~(\ref{eq.onsager1}) and (\ref{eq.onsager2}), 
we find that the maximum power is realized at $X_1^*=-\frac{L_{12}X_2}{2L_{11}}$ from $\frac{\partial \dot{W}}{\partial X_1}=0$. Its efficiency $\eta^*$ is given as
\begin{eqnarray}
\eta^*=\frac{q^2}{2-q^2}\frac{\Delta T^{\rm R}}{2T^{\rm R}},\label{eq.eta_maxp_q}
\end{eqnarray}
which is a monotonically increasing function of $|q|$, where the coupling strength $q$ is defined by
\begin{eqnarray}
q\equiv \frac{L_{12}}{\sqrt{L_{11}L_{22}}}.\label{eq.def_q}
\end{eqnarray}
From the non-negativity of the entropy production rate $\dot{\sigma}=J_1X_1+J_2X_2$, the Onsager coefficients $L_{ij}$'s should satisfy $L_{11}\ge 0$, $L_{22}\ge 0$, and $L_{11}L_{22}-L_{12}^2\ge 0$, and they impose the following constraint on $q$:
\begin{eqnarray}
|q|\le 1,
\end{eqnarray}
where the equality is known as the tight-coupling (no heat-leakage) condition~\cite{kc,VB}. Under this tight-coupling condition, 
$\eta^*$ in Eq.~(\ref{eq.eta_maxp_q}) attains the upper bound given by the CA efficiency $\eta_{\rm CA}$ as in Eq.~(\ref{eq.ca.bound}).
An essential point of the derivation of the formula Eq.~(\ref{eq.eta_maxp_q}) is that 
the non-zero cross-coefficient $L_{12}$ plays an important role in $\eta^*$ in Eq.~(\ref{eq.eta_maxp_q}), which is clear from the definition Eq.~(\ref{eq.def_q}).

Returning to our original problem, from Eq.~(\ref{eq.jq_fourier}), we formally obtain the following ``Onsager coefficients" under our choice of the thermodynamic fluxes $J_{Q_i}$ and forces $X_{T_i}$:
\begin{eqnarray}
L_{ij}=
\begin{pmatrix}
a_h\kappa_h{T^{\rm R}}^2 & 0\\
0 & a_c \kappa_c {T^{\rm R}}^2
\end{pmatrix},\label{eq.onsager_matrix1}
\end{eqnarray}
where there are no nondiagonal elements. 
This contrasts to the formulation using Eqs.~(\ref{eq.onsager1}) and (\ref{eq.onsager2}) where the cross-terms play an important role in the heat-flux conversion into power~\cite{VB}. 
Eq.~(\ref{eq.onsager_matrix1}) is natural if the entropy production originating from the heat transfer between the working substance and the heat reservoir in each isothermal process is independent of each other. 
However, as seen from Eqs.~(\ref{eq.xth_approx}) and (\ref{eq.xtc_approx}), $X_{T_i}$'s are not independent of each other, unlike $X_i$'s in Eq.~(\ref{eq.entropy_vb}), 
but $X_{T_i}$'s and $X_i$'s should be related with each other by a variable change.
Moreover, how the external force $F$ in the definition of $X_1=F/T^{\rm R}$ against the heat engine~\cite{VB} is related to the thermodynamic variable of the working substance 
is not obvious in our local equilibrium description of the finite-time Carnot cycle where the efficiency and the power is expressed by them as in Eqs.~(\ref{eq.endo_carnot}) and (\ref{eq.power2}).
To elucidate these points, we restate our expression of the entropy production rate Eq.~(\ref{eq.entropy_product}) with $X_{T_i}$'s 
using independent thermodynamic forces as (see Fig.~\ref{engine} (b))
\begin{eqnarray}
\dot{\sigma}=\frac{1}{t_{\rm cyc}}\oint d{s}
=\sum_i \left(\frac{1}{T_i^{\rm R}}-\frac{1}{T_i}\right)\frac{\Delta U_i^{\rm R}}{t_{\rm cyc}}
=\left(\frac{1}{T_c^{\rm R}}-\frac{1}{T_h^{\rm R}}\right)\frac{Q_h}{t_{\rm cyc}}-\frac{\Delta T}{T_c^{\rm R}}\frac{\Delta S}{t_{\rm cyc}}
\equiv J_{Q_h} Y_{T^{\rm R}}+J_S Y_T.\label{eq.JSJQ}
\end{eqnarray}
We can make this restatement by using 
the endoreversibility condition Eq.~(\ref{eq.endoreversibility}),
the first law of the thermodynamics $W=Q_h+Q_c$ for one cycle, 
and Eq.~(\ref{eq.power2}), where we defined the heat flux from the hot heat reservoir 
as a new thermodynamic flux and its conjugate new thermodynamic force as
\begin{eqnarray}
J_{Q_h}=\frac{Q_h}{t_{\rm cyc}}=a_hT_h\frac{dS_h}{dt}, \ Y_{T^{\rm R}} \equiv \frac{1}{T_c^{\rm R}}-\frac{1}{T_h^{\rm R}}.\label{eq.JQYTR}
\end{eqnarray}
In addition, we defined the entropy flux as another new thermodynamic flux and its conjugate new thermodynamic force as
\begin{eqnarray}
J_S\equiv \frac{\Delta S}{t_{\rm cyc}}=a_h\frac{dS_h}{dt}, \  Y_T\equiv -\frac{\Delta T}{T_c^{\rm R}}.\label{eq.JSYT}
\end{eqnarray}
In this way, all the thermodynamic fluxes and forces are expressed in terms of the combination of the thermodynamic variables of 
the working substance and the heat reservoirs owing to the local equilibrium assumption.
From Eq.~(\ref{eq.JSYT}), in particular, the new thermodynamic force $Y_T$ is proportional to
the temperature difference of the working substance between the isothermal processes (Fig.~\ref{engine} (b)), which has never been proposed so far. 
The external force $F$ in the linear irreversible thermodynamics framework in Eq.~(\ref{eq.ex_force}) by which the power is maximized  
is abstract in the case of the finite-time Carnot cycle. But our definition $Y_T$ in Eq.~(\ref{eq.JSYT}) gives us a more intuitive picture of ``force" applied to the finite-time Carnot cycle, which can be used as a parameter for maximization of the power. This is also consistent with the Curzon and Ahlborn's original idea that reminded them of the maximization of the power in their finite-time Carnot cycle~\cite{CA}: the power becomes zero in the quasistatic limit $\Delta T=\Delta T^{\rm R}$ and also in the other extreme situation $\Delta T=0$ where the heat flux from the hot heat reservoir is just transferred to the cold heat reservoir without generating any work. Then the power must take a maximum with an intermediate $\Delta T$ between these two zeros.

In the linear response regime, these new thermodynamic fluxes and forces are approximated as 
\begin{eqnarray}
J_{Q_h}\simeq a_hT^{\rm R}\frac{dS_h}{dt}, \ Y_{T^{\rm R}} \simeq \frac{\Delta T^{\rm R}}{{T^{\rm R}}^2}, J_S=\frac{\Delta S}{t_{\rm cyc}}=a_h\frac{dS_h}{dt}, \  Y_T\simeq -\frac{\Delta T}{T^{\rm R}}.\label{eq.new_thermo_force_lin}
\end{eqnarray}
We note that $Y_T$ ($Y_{T^{\rm R}}$) corresponds to $X_1$ ($X_2$), and $J_S$ ($J_{Q_h}$) does to $J_1$ ($J_2$) in the definition of the general formulation in Eq.~(\ref{eq.entropy_vb}), respectively.
From Eq.~(\ref{eq.new_thermo_force_lin}), we immediately notice that $J_{Q_h}$ and $J_S$ are in proportion to each other:
\begin{eqnarray}
J_{Q_h}=T^{\rm R}J_S.\label{eq.JQJS}
\end{eqnarray}
In fact, the proportionality between the two thermodynamic fluxes in Eq.~(\ref{eq.JQJS}) indirectly implies the tight-coupling condition of this system $|q|=1$~\cite{WT}, 
because we can easily show  from Eqs.~(\ref{eq.onsager1}) and (\ref{eq.onsager2}) the relation $J_2=\frac{L_{21}}{L_{11}}J_1+L_{22}(1-q^2)X_2$ between the two thermodynamic fluxes.
However, to understand this relationship more directly and precisely, we express the present system by the following Onsager relations using the new thermodynamic fluxes and forces:
\begin{eqnarray}
J_{S}&&=\tilde{L}_{TT}Y_T+\tilde{L}_{TT^{\rm R}}Y_{T^{\rm R}},\label{eq.onsager_JS}\\
J_{Q_h}&&=\tilde{L}_{T^{\rm R}T}Y_T+\tilde{L}_{T^{\rm R}T^{\rm R}}Y_{T^{\rm R}}.\label{eq.onsager_JQ}
\end{eqnarray}
To obtain the new Onsager coefficients from the previous coefficients in Eq.~(\ref{eq.onsager_matrix1}), 
we relate the thermodynamic forces $X_{T_i}$ ($i=h,c$) and $Y_m$ ($m=T, T^{\rm R}$) by using Eqs.~(\ref{eq.xth_approx}), (\ref{eq.xtc_approx}), and (\ref{eq.new_thermo_force_lin}) as follows:
\begin{eqnarray}
\begin{pmatrix}
X_{T_h}\\
X_{T_c}
\end{pmatrix}
=\frac{X_T}{a_h\kappa_h+a_c\kappa_c}
\begin{pmatrix}
{a_c\kappa_c}\\
-{a_h\kappa_h}
\end{pmatrix}
=\frac{1}{(a_h\kappa_h+a_c\kappa_c)T^{\rm R}}
\begin{pmatrix}
{a_c\kappa_c} & a_c\kappa_c T^{\rm R}\\
-{a_h\kappa_h} & -{a_h\kappa_h} T^{\rm R}
\end{pmatrix}
\begin{pmatrix}
Y_T \\
Y_{T^{\rm R}}
\end{pmatrix}\label{eq.x_y}
.
\end{eqnarray}
Rewriting Eq.~(\ref{eq.x_y}) as $X_{T_i}\equiv F_{im}Y_m$ in Einstein notation, 
we obtain the new Onsager matrix 
\begin{eqnarray}
\tilde{L}_{mn}=
\frac{a_h \kappa_h a_c\kappa_c}{a_h\kappa_h+a_c\kappa_c}
\begin{pmatrix}
1 & T^{\rm R} \\
{T^{\rm R}} & {T^{\rm R}}^2
\end{pmatrix},\label{eq.onsager_matrix_2}
\end{eqnarray}
from the relation $\tilde{L}_{mn}=F_{mi}^{\rm T}L_{ij}F_{jn}$ that conserves the 
entropy production rate as $\dot{\sigma}=L_{ij}X_{T_i}X_{T_j}=\tilde{L}_{mn}Y_mY_n$.
Alternatively, we can directly obtain the Onsager coefficients $\tilde{L}_{T^{\rm R}T}$ and $\tilde{L}_{T^{\rm R}T^{\rm R}}$
from the expression of
\begin{eqnarray}
J_{Q_h}=a_h\kappa_h{T^{\rm R}}^2X_{T_h}=\frac{a_h\kappa_h a_c\kappa_c}{a_h\kappa_h+a_c\kappa_c}{T^{\rm R}}^2X_T=\frac{a_h \kappa_h a_c\kappa_c}{a_h\kappa_h+a_c\kappa_c}T^{\rm R}Y_T+\frac{a_h \kappa_h a_c\kappa_c}{a_h\kappa_h+a_c\kappa_c}{T^{\rm R}}^2Y_{T^{\rm R}},
\end{eqnarray}
which is obtained using Eqs.~(\ref{eq.xth_approx}) and (\ref{eq.x_y}), and we can also obtain $\tilde{L}_{TT}$ and $\tilde{L}_{TT^{\rm R}}$ from Eq.~(\ref{eq.JQJS}):
\begin{eqnarray}{
J_{S}=\frac{J_{Q_h}}{T^{\rm R}}=a_h\kappa_h{T^{\rm R}}X_{T_h}=\frac{a_h\kappa_h a_c\kappa_c}{a_h\kappa_h+a_c\kappa_c}{T^{\rm R}}X_T=\frac{a_h \kappa_h a_c\kappa_c}{a_h\kappa_h+a_c\kappa_c}Y_T+\frac{a_h \kappa_h a_c\kappa_c}{a_h\kappa_h+a_c\kappa_c}{T^{\rm R}}Y_{T^{\rm R}}}\label{eq.JS2}.
\end{eqnarray}
From Eq.~(\ref{eq.onsager_matrix_2}), it is straightforward to confirm that the Onsager reciprocity and the tight-coupling (no heat-leakage) condition are fulfilled:
\begin{eqnarray}
\tilde{L}_{TT^{\rm R}}=\tilde{L}_{T^{\rm R}T}, \ \ q=\frac{\tilde{L}_{TT^{\rm R}}}{\sqrt{\tilde{L}_{T^{\rm R}T^{\rm R}}\tilde{L}_{TT}}}=1.
\end{eqnarray}
This implies that our model attains the upper bound (the CA efficiency) in Eq.~(\ref{eq.ca.bound}) corresponding to $|q|=1$ in Eq.~(\ref{eq.eta_maxp_q}), from the viewpoint of the linear irreversible thermodynamics framework. 
For completeness, we explicitly confirm that the maximization of the power by $Y_T$ leads to the CA efficiency by using the Onsager relations along with the general formulation introduced in the beginning of this subsection.
Let us compare the two ways of maximization of the power by $X_T$ as in Sec.~II B and by $Y_T$. 
As we have seen in Eq.~(\ref{eq.x_y}), the reduced thermodynamic force $X_T$ introduced in Eq.~(\ref{eq.reduced_force}) is linearly connected by $Y_T$ and $Y_{T^{\rm R}}$.
By substituting Eq.~(\ref{eq.x_y}) into Eq.~(\ref{eq.pow_xt}), we can express the power $\dot{W}$ in Eq.~(\ref{eq.pow_xt}) by using $Y_T$ and $Y_{T^{\rm R}}$ instead of using $X_T$ as
\begin{eqnarray}
\dot{W}=-\frac{a_h \kappa_h a_c\kappa_c}{a_h\kappa_h+a_c\kappa_c}T^{\rm R}Y_T^2-\frac{a_h \kappa_h a_c \kappa_c}{a_h\kappa_h+a_c\kappa_c}{T^{\rm R}}^2Y_{T^{\rm R}}Y_T,\label{eq.pow_yt}
\end{eqnarray}
which can also be obtained from the expression $\dot{W}=-J_SY_TT^{\rm R}$ in Eq.~(\ref{eq.ex_force}) and Eq.~(\ref{eq.onsager_JS}).
We find that the quasistatic limit $X_T^{\rm qs}=0$ corresponds to $Y_T^{\rm qs}=-T^{\rm R}Y_{T^{\rm R}}$ from $\dot{W}=0$ by using Eq.~(\ref{eq.pow_yt}).
We also find that the maximum power point $X_T^*=\frac{\Delta T^{\rm R}}{2{T^{\rm R}}^2}$ in Eq.~(\ref{eq.maxpow_condition}) also corresponds to $Y_T^*=-\frac{T^{\rm R}}{2}Y_{T^{\rm R}}$ from $\frac{\partial \dot{W}}{\partial Y_T}=0$ by using Eq.~(\ref{eq.pow_yt}), from which we can obtain the efficiency $\eta^*=-J_S^*Y_T^*T^{\rm R}/J_{Q_h}^*$ 
by using Eqs.~(\ref{eq.onsager_JS}) and (\ref{eq.onsager_JQ}) as
\begin{eqnarray}
\eta^*=\frac{\Delta T^{\rm R}}{2T^{\rm R}}=\eta_{\rm CA}+O\left({\Delta T^{\rm R}}^2\right),\label{eq.ca_proved}
\end{eqnarray}
Therefore, for our endoreversible heat engine model based on the local equilibrium assumption, we conclude that the efficiency at maximum power 
attains the upper bound in Eq.~(\ref{eq.ca.bound}) from a viewpoint of the linear irreversible thermodynamics framework using the Onsager relations~\cite{VB}.
We note that the above derivation is quite general because it does not rely on any particular working substance or thermal conductance.

 \section{Discussion and Summary}
We discuss a few aspects related to our formulation in Sec.~II.

First, we note that the endoreversible power as determined by the product of the heat flux and the endoreversible efficiency in Eq.~(\ref{eq.pow_xt}) in Sec.~II B should also be determined by the first law of the thermodynamics 
$\dot{W}=J_{Q_h}+J_{Q_c}$ for one cycle.
To this end, we need to consider $J_{Q_i}$ with higher-order corrections of $\Delta T$ and $\Delta T^{\rm R}$, while we have considered only the 
lowest order in Sec.~II B. 
We utilize the expressions $J_{Q_h}=T_hJ_S$ and $J_{Q_c}=-T_cJ_S$, which are obtained from Eqs.~(\ref{eq.Qc}), (\ref{eq.entropy_change}), (\ref{eq.def_flux_force}), and (\ref{eq.JSYT}). 
Then, by using Eqs.~(\ref{eq.Th}), (\ref{eq.Tc}), and (\ref{eq.JS2}), we directly obtain the explicit form of $J_{Q_i}$ with the nonlinear terms as
\begin{eqnarray}
&&J_{Q_h}=\frac{a_h\kappa_ha_c \kappa_c}{a_h\kappa_h+a_c\kappa_c} {T^{\rm R}}^2 X_T+\frac{a_h\kappa_ha_c\kappa_c}{2(a_h\kappa_h+a_c\kappa_c)}T^{\rm R}\Delta T^{\rm R}X_T-\frac{a_h\kappa_ha_c^2\kappa_c^2}{(a_h\kappa_h+a_c\kappa_c)^2}{T^{\rm R}}^3X_T^2
,\label{eq.JQ_linear}\\
&&J_{Q_c}=-\frac{a_h\kappa_ha_c \kappa_c}{a_h\kappa_h+a_c\kappa_c} {T^{\rm R}}^2 X_T+\frac{a_h\kappa_ha_c\kappa_c}{2(a_h\kappa_h+a_c\kappa_c)}T^{\rm R}\Delta T^{\rm R}X_T-\frac{a_h^2\kappa_h^2a_c\kappa_c}{(a_h\kappa_h+a_c\kappa_c)^2}{T^{\rm R}}^3X_T^2
,\label{eq.JQ_linear2}
\end{eqnarray}
from which we can confirm $\dot{W}=J_{Q_h}+J_{Q_c}$ by using Eq.~(\ref{eq.pow_xt}).

Second, we point out that an extension of our formulation introduced in Sec.~II A to the general case where the energy flux may not be constant along the cycle may be an interesting challenge. 
This would be more clarified by considering under what conditions the assumption of the constant energy flux strictly holds in a specific example:  
let us consider the following first law of thermodynamics Eq.~(\ref{eq.JQ}) for the $3$-dimensional ideal gas during an isothermal process:
\begin{eqnarray}
\frac{3}{2}Nk_{\rm B}\frac{dT_i}{dt}=\kappa_i (T_i^{\rm R}-T_i)-\frac{Nk_{\rm B}T_i}{V_i}\frac{dV_i}{dt},\label{eq.1st_law_idealgas}
\end{eqnarray}
where we used the internal energy $U_i=\frac{3}{2}Nk_{\rm B}T_i$ and the equation of state $P_i=\frac{Nk_{\rm B}T_i}{V_i}$ for the ideal gas, where we denote by $N$ and $k_{\rm B}$ the particle number and the Boltzmann constant, respectively.
For each term in Eq.~(\ref{eq.1st_law_idealgas}) to be constant, 
$T_i={\rm const}$ solution to Eq.~(\ref{eq.1st_law_idealgas}) can be 
realized only under a specific protocol $V_i(t)$ that satisfies $\frac{1}{V_i}\frac{dV_i}{dt}={\rm const}$.
Therefore even in this simplest case of the ideal gas, our assumption of the constant energy flux introduced in Sec.~II A 
(as also assumed in the original CA model) seems too restricted. In addition, although we assumed that the thermal conductance $\kappa_i$ is a 
constant in the present model (as also assumed in the original CA model), 
it can also depend on the thermodynamic variables such as $T_i$ and $V_i$~\cite{IO2009,SS}, and hence on time through them in general.
Since actual heat engines may consist of complicated combinations of these factors, it is difficult to specify generic conditions for arbitrary working substances and thermal conductances
under which the assumption of the constant energy flux holds in a strict manner.
Instead, for more natural description of the endoreversible finite-time Carnot cycle under more generic conditions, 
we are motivated to extend our formulation introduced in Sec.~II A to non-constant energy flux cases that include the constant energy flux case as a special case. 
We are also in progress toward this goal from the linear irreversible thermodynamics point of view, which will be presented elsewhere~\cite{IOpreparation}.
However, we stress that our present formulation of the constant energy flux case would serve as the ``zeroth approximation" or a starting point
for analyzing the performance of such actual heat engines. 
Indeed, this is also supported by the fact that observed data of the efficiency of actual heat engines are distributed around the CA efficiency~\cite{EKLVB}, 
which implies that they may be regarded as more or less variants of the constant energy flux case.

In the present study, we formulated an endoreversible finite-time Carnot cycle model based on the assumptions of local equilibrium and constant energy flux.
In our framework, the power and the efficiency are expressed in terms of the thermodynamic variables of the working substance.
From the analysis of the entropy production rate caused by the heat transfer in each isothermal process, 
we identified the thermodynamic flux and force in each isothermal process, which are related by the Fourier law.
By applying the endoreversible condition to the linear response regime, we found that those thermodynamic forces are not independent, and identified
the reduced thermodynamic force and its conjugate thermodynamic flux, which are connected by the linear relation different from the Onsager relations.
We calculated the efficiency at maximum power by using this linear relation, and obtained the Curzon-Ahlborn efficiency.
We also elucidated that by suitable change of the variables, the linear relation between the thermodynamic flux and force in our framework can be rewritten into the form of the 
Onsager relations, where the novel thermodynamic force that is proportional to the temperature difference of the working substance between the isothermal processes is introduced.
Then we directly confirmed that our model satisfies the tight-coupling condition that ensures the Curzon-Ahlborn efficiency as is the upper bound in the linear irreversible thermodynamics framework. We stress that our framework is quite universal because it only assumes that the working substance is in a local equilibrium state specified by a unique combination of thermodynamic variables at any instant along the cycle.
We expect that our study unifies recent development of the theories of heat engines 
based on the universal nonequilibrium thermodynamics framework and the more phenomenological finite-time thermodynamics approach that was developed for application to real power plants and heat devices.

\begin{acknowledgements} 
The authors are grateful to anonymous referees for helpful comments on the manuscript.
Y. I. acknowledges the financial support from a Grant-in-Aid for JSPS Fellows (Grant No. 25-9748).
\end{acknowledgements}

\end{document}